\def\VersionLong{}
\def\VersionFinal{}
\def\VersionArxiv{}
\ifdefined\VersionLong%
	\newcommand{\LongVersion}[1]{\#1}
	\newcommand{\ShortVersion}[1]{}
\else
	\newcommand{\LongVersion}[1]{}
	\newcommand{\ShortVersion}[1]{#1}
\fi

\ifdefined\VersionArxiv%
\documentclass{scrartcl}
\else
\documentclass[runningheads,a4paper,10pt,envcountsame]{llncs}
\fi

\usepackage{comment}
\ifdefined\VersionLong%
        \includecomment{LongVersionBlock}
        \excludecomment{ShortVersionBlock}
\else
        \excludecomment{LongVersionBlock}
        \includecomment{ShortVersionBlock}
\fi

\usepackage[ruled,lined,linesnumbered,noend]{algorithm2e}
\DontPrintSemicolon{}
\SetKwInOut{Input}{Input}
\SetKwInOut{Output}{Output}
\SetKw{KwRemove}{remove}
\SetKw{KwPush}{add}
\SetKw{KwPop}{pop}
\SetKw{KwFrom}{from}
\SetKw{KwTo}{to}
\SetKw{KwCompute}{compute}
\SetKw{KwReturn}{return}
\SetKw{KwBreak}{break}
\SetKw{KwPick}{pick}
\SetKw{KwLet}{let}
\SetKw{KwBe}{be}
\SetKw{KwSplit}{split}
\SetKw{KwInto}{into}
\SetKw{KwFind}{find}
\SetKw{KwFeed}{feed}
\SetKwProg{Fn}{Function}{:}{}
\SetAlgorithmName{Algorithm}{algorithm}{List of Algorithms}

\usepackage[svgnames,table]{xcolor}

\usepackage{graphicx}
\usepackage{stmaryrd} %
\usepackage{amssymb}
\usepackage{amsmath}
\usepackage{siunitx}
\usepackage{fontawesome5}
\usepackage[pdf,singlefile]{graphviz}
\usepackage{amsfonts}
\usepackage{tikz}
\usepackage{booktabs}
\usetikzlibrary{graphs}
\usetikzlibrary{automata,positioning,arrows.meta,decorations.pathmorphing,decorations.pathreplacing,decorations.shapes}
\usepackage{multirow}
\usepackage{array} 
\usepackage{float}
\usepackage{listings}
\AtBeginDocument{}

\lstdefinestyle{json}{
    backgroundcolor=\color{white},   %
    basicstyle={\footnotesize\ttfamily},
    numbers=none,                    %
    showstringspaces=false,
    breaklines=true,
    frame=lines,
    captionpos=b,                    %
    keywordstyle=[1]\color{red!70!black}\bfseries,       %
    morekeywords=[1]{time, package_name, package_tag, level, ts, msg},            %
}

\lstdefinestyle{log}{
	backgroundcolor=\color{white},   %
	basicstyle={\scriptsize\ttfamily},        %
	breakatwhitespace=false,         %
	breaklines=true,                 %
	captionpos=b,                    %
	commentstyle=\color{green},    %
	deletekeywords={...},            %
	escapeinside={\%*}{*)},          %
	extendedchars=true,              %
	frame=lines,	                   %
	keepspaces=true,                 %
	keywordstyle=[1]\color{red!70!black}\bfseries,       %
	morekeywords=[1]{create, fetch},            %
        otherkeywords={auth-backend, auth-example, auth-frontend},
        morekeywords=[2]{auth-backend, auth-example, auth-frontend},
        keywordstyle=[2]\color{blue},
	numbers=left,                    %
	numbersep=0pt,                   %
	numberstyle=\tiny\color{gray}, %
	rulecolor=\color{black},         %
	showspaces=false,                %
	showstringspaces=false,          %
	showtabs=false,                  %
	stepnumber=1,                    %
	stringstyle=\color{mauve},     %
	tabsize=2,	                   %
}

\lstdefinestyle{symon}{
	backgroundcolor=\color{white},   %
	basicstyle={\scriptsize\ttfamily},        %
	breakatwhitespace=false,         %
	breaklines=true,                 %
	captionpos=b,                    %
	commentstyle=\color{green},    %
	deletekeywords={...},            %
	escapeinside={\%*}{*)},          %
	extendedchars=true,              %
	frame=lines,	                   %
	keepspaces=true,                 %
	keywordstyle=[1]\color{red!70!black}\bfseries,       %
	morekeywords=[1]{create, fetch},            %
        otherkeywords={auth-backend, auth-example, auth-frontend},
        morekeywords=[2]{var, signature, expr, zero_or_more, within, one_of, or},
        keywordstyle=[2]\color{blue},
        morekeywords=[3]{ignore_any, ignore_irrelevant, correct, failed},
        keywordstyle=[3]\color{teal},
        numbers=left,                    %
	numbersep=0pt,                   %
	numberstyle=\tiny\color{gray}, %
	rulecolor=\color{black},         %
	showspaces=false,                %
	showstringspaces=false,          %
	showtabs=false,                  %
	stepnumber=1,                    %
	stringstyle=\color{mauve},     %
	tabsize=2,	                   %
}

\lstdefinestyle{symon_result}{
	backgroundcolor=\color{white},   %
	basicstyle={\scriptsize\ttfamily},        %
	breakatwhitespace=false,         %
	breaklines=true,                 %
	captionpos=b,                    %
	commentstyle=\color{green},    %
	deletekeywords={...},            %
	escapeinside={\%*}{*)},          %
	extendedchars=true,              %
	frame=lines,	                   %
	keepspaces=true,                 %
        otherkeywords={@, time-point},
        morekeywords=[1]{@, time-point},
        keywordstyle=[1]\color{blue},
	numbers=left,                    %
	numbersep=0pt,                   %
	numberstyle=\tiny\color{gray}, %
	rulecolor=\color{black},         %
	showspaces=false,                %
	showstringspaces=false,          %
	showtabs=false,                  %
	stepnumber=1,                    %
	stringstyle=\color{mauve},     %
	tabsize=2,	                   %
}

\usepackage{colortbl}
\definecolor{darkblue}{rgb}{0.0,0.0,0.6}
\definecolor{darkgreen}{rgb}{0, 0.5, 0}
\definecolor{darkpurple}{rgb}{0.7, 0, 0.7}
\definecolor{darkblue}{rgb}{0, 0, 0.7}

\usepackage[
	colorlinks=true,
	citecolor=darkgreen,
	linkcolor=darkblue,
	urlcolor=darkpurple,
]{hyperref}

\usepackage[capitalise,english,nameinlink]{cleveref} %
\crefalias{AlgoLine}{line}
\crefname{line}{\text{line}}{\text{lines}} %
\crefname{item}{\text{item}}{\text{items}} %
\crefname{example}{\text{Example}}{\text{Examples}} %
\crefname{assumption}{\text{Assumption}}{\text{Assumptions}} %
\crefname{algorithm}{\text{Algorithm}}{\text{Algorithms}}

\ifdefined\VersionWithComments%
	\usepackage{draftwatermark}
	\SetWatermarkText{draft}
 	\SetWatermarkScale{15}
	\SetWatermarkColor[gray]{0.9}
\fi

\ifdefined\VersionWithComments%
	\usepackage[colorinlistoftodos,textsize=footnotesize]{todonotes}
\else
	\usepackage[disable]{todonotes}
\fi

\ifdefined\VersionFinal%
\else
	\usepackage[pagewise]{lineno} %
	\linenumbers%
	
\fi

\tikzstyle{defproblem} = [
 draw=black,
 fill=cyan!10,
 text=black,
 line width=0.5pt,
  text width = \linewidth - 1.6 ex - 1pt,
  inner sep = 0.8 ex,
  rounded corners=4pt]
\tikzstyle{rqanswer} = [
 draw=black,
 fill=gray!30,
 text=black,
 line width=0.5pt,
  text width = \linewidth - 1.6 ex - 1pt,
  inner sep = 0.8 ex,
  rounded corners=4pt]

\usepackage{subcaption}

\usepackage{paralist} %

\usepackage{booktabs}

\newcommand{\word}[1][]{w#1}

\newcommand{\Autom}{\mathcal{A}}

\ifdefined\VersionWithComments%
 	\definecolor{colorok}{RGB}{80,80,150}
\else
	\definecolor{colorok}{RGB}{0,0,0}
\fi

\makeatletter
\def\orcidID#1{\smash{\href{https://orcid.org/#1}{\protect\raisebox{-1.25pt}{\protect\includegraphics{ORCID_Color.eps}}}}}
\makeatother

\makeatletter
\AtBeginDocument{%
  \@ifpackageloaded{hyperref}
  {\def\@doi#1{\href{https://doi.org/#1}
      {\ttfamily https://doi.org/#1}\egroup}}
  {\def\@doi#1{\ttfamily https://doi.org/#1\egroup}}
  \def\doi{\bgroup\catcode`\_=12\relax\@doi}}
\makeatother

\ifdefined\VersionArxiv%
\usepackage{authblk}
\usepackage{orcidlink}
\fi

\begin{document}
\title{A Case Study on Runtime Verification of a Continuous Deployment Process\ifdefined\VersionArxiv\thanks{%
This is the author version of an extended abstract for the presentation at RVCase 2025 (\url{https://seanmk.com/rvcase/}).%
}\fi}
\ifdefined\VersionArxiv%
\author[1]{Shoma Ansai}
\author[1,2]{Masaki Waga\orcidlink{0000-0001-9360-7490}}
\date{\vspace{-5ex}}
\else
\author{%
Shoma Ansai\inst{1}\and
Masaki Waga\orcidID{0000-0001-9360-7490}\inst{1,2}%
}
\fi
\ifdefined\VersionArxiv%
\else
\authorrunning{%
S.\ Ansai \& M.\ Waga
}
\fi
\ifdefined\VersionArxiv%
\affil[1]{Graduate School of Informatics, Kyoto University, Kyoto, Japan}
\affil[2]{National Institute of Informatics, Tokyo, Japan}
\else
\institute{%
Graduate School of Informatics, Kyoto University, Kyoto, Japan
\and
National Institute of Informatics, Tokyo, Japan
}
\fi
\maketitle              %
\ifdefined\VersionArxiv
\else
\pagestyle{plain}
\fi
\begin{abstract}
We report our experience in applying runtime monitoring to a FluxCD-based continuous deployment (CD) process. Our target system consists of GitHub Actions, GitHub Container Registry (GHCR), FluxCD, and an application running on Kubernetes. We monitored its logs using SyMon. In our setting, we regard a deployment update as detected when FluxCD's polling log resolves the latest image tag. Through the case study, we found that FluxCD did not always detect a new image within five minutes after it was pushed to GHCR, whereas it always did so within ten minutes in the collected logs. Moreover, our results show that SyMon is fast enough for near-real-time monitoring in our setting.

\ifdefined\VersionArxiv%
\else
\fi
\end{abstract}

\section{Introduction}

Modern web applications are usually deployed using containers.
A container is a lightweight virtual environment. Deploying an application in a container enables seamless server setup because the container includes all required libraries, runtime environments, and configurations. Application developers build a container image and deploy it to production as a container.
Meanwhile, continuous deployment (CD) is gaining attention. CD is a software development strategy in which code changes are automatically deployed to the production environment. It enables rapid and safe deployments. CD can be divided into the following two processes.

\begin{enumerate}
    \item \textbf{Building Phase}: Build an image from the source code.
    \item \textbf{Deployment Phase}: Create a container based on the built image.
\end{enumerate}

There are three main ways to trigger the second step after the first completes.

\begin{itemize}
    \item Manually start the deployment.
    \item Trigger the deployment immediately after the image build is completed.
    \item Periodically poll the container image registry, and deploy the image when a new version is found.
\end{itemize}

Tools such as FluxCD~\cite{fluxcd} and ArgoCD~\cite{argocd}, which run in the Kubernetes~\cite{kubernetes} environment, operate based on the third approach. However, this method presents a challenge. Because the deployment process is triggered asynchronously after the image is built, it becomes important to confirm that the CD controller promptly detects newly built images. Delays or failures in detecting or deploying a new image can cause serious problems. For example, if a security patch is built into a new image but the update is delayed due to an unforeseen error, the vulnerability may remain in production for an extended period, potentially leading to critical security breaches.

In this paper, we report our experience in applying runtime verification to a FluxCD-based CD process to detect unexpected behavior. Specifically, we monitored its logs using SyMon~\cite{DBLP:conf/cav/WagaAH19}. In the collected logs, FluxCD did not always detect a new image within five minutes after it was pushed to GHCR, whereas it always did so within ten minutes. Moreover, our results show that SyMon is fast enough for near-real-time monitoring in our setting.
\section{Related Work}

Continuous Integration (CI) mechanisms are widely used to verify deployed applications~\cite{netflix-testing,7884954,uber-testing}, but to the best of our knowledge, there are no reported examples of runtime verification of the CD system itself.

It is common practice to use application metrics to monitor whether the system is running properly~\cite{boniol2024divetimeseriesanomalydetection,trivago-monitoring,netflix-monitoring,uber-monitoring,zhang2022movingmetricdetectionalerting}. However, we have not found examples that monitor whether a CD system satisfies specific time constraints based on its output logs.

A method for statically verifying Kubernetes behavior has been proposed~\cite{298551}. However, this method only verifies the behavior of deploying Kubernetes pods and cannot be used to verify the behavior of the entire CD system. A mechanism for monitoring Kubernetes events in real time and detecting abnormal behavior has also been proposed~\cite{VarunKumar2024-VAREKM}, but we have not found examples that monitor whether the system satisfies specific time constraints.

\section{Symbolic Monitoring with SyMon}

SyMon~\cite{DBLP:conf/cav/WagaAH19} is a tool for symbolic monitoring of \emph{timed data words} against \emph{parametric timed data automata (PTDAs)}.
A timed data word is a \emph{timed word}~\cite{DBLP:journals/tcs/AlurD94} equipped with infinite domain data (e.g., strings and numbers) that represent, for instance, identifiers or sensed values.
Specifically, a timed data word is a sequence of (finite domain) events associated with infinite domain data and timestamps.
\cref{code:log-for-symon} shows a concrete timed data word that SyMon can handle, where the set of events is $\{\text{create}, \text{fetch}\}$.
Each line in \cref{code:log-for-symon} represents an event associated with two string values representing the package name and package tag (second and third columns), and a timestamp (fourth column).

\begin{lstlisting}[caption={A log compatible with SyMon. Each line represents an event equipped with infinite domain data and a timestamp. The first column shows the event, the second column shows the package name, the third column shows the package tag, and the fourth column shows the timestamp.},label=code:log-for-symon,style=log,float]
create auth-backend stg-7c03f5241c93d6e77bb132d8ea9ffe9e59e7b62d-1445 171982
fetch auth-example stg-379cca639565f93fe2485c6f443b1d5b45285534-1441 172084
fetch auth-example stg-379cca639565f93fe2485c6f443b1d5b45285534-1441 172085
create auth-frontend stg-7c03f5241c93d6e77bb132d8ea9ffe9e59e7b62d-1445 172140
fetch auth-frontend stg-7c03f5241c93d6e77bb132d8ea9ffe9e59e7b62d-1445 172146
\end{lstlisting}

PTDAs are a generalization of \emph{parametric timed automata}~\cite{DBLP:conf/stoc/AlurHV93} (which extend timed automata~\cite{DBLP:journals/tcs/AlurD94} with parameters in timing constraints) to handle infinite domain data.
Informally, a PTDA is an NFA equipped with \emph{clock variables} and \emph{data variables} to represent constraints on the time gap between events and the infinite domain data on events, respectively.
More specifically, each transition of a PTDA is labeled with parametric constraints and updates on clock and data variables.
See~\cite{DBLP:conf/cav/WagaAH19} for the details of PTDAs.

The semantics of PTDAs is defined with respect to valuations of the timing and data parameters in the constraints.
Namely, the parametric constraints in a PTDA are instantiated with the parameter valuations, and its language is defined based on the instantiated concrete constraints.
Given a timed data word $\word$ and a PTDA $\Autom$, SyMon returns the prefixes of $\word$ accepted by $\Autom$ along with the corresponding set of parameter valuations $\eta$.
Thanks to the parameters, one can formulate a PTDA $\Autom$ with unknown values, and SyMon can detect the acceptance of prefixes of $\word$ along with the concrete instance of the unknown values.
For example, one can represent a violation of certain requirements \emph{for some} identifier of interest as a PTDA and use SyMon to detect its violation and obtain the concrete identifier for each detected violation.

\begin{lstlisting}[caption=SyMon's specification,label=code:symon-specification,style=symon,float]
#!/usr/local/bin/symon -dnf
var {
    current_name: string;
    current_tag: string;
}
signature create {
    name: string;
    tag: string;
}
signature fetch {
    name: string;
    tag: string;
}
expr ignore_any {
    zero_or_more {
        one_of {
            create(name, tag)
        } or {
            fetch(name, tag)
        }
    }
}
expr ignore_irrelevant {
    zero_or_more {
        one_of {
            create(name, tag | name != current_name || tag != current_tag)
        } or {
            fetch(name, tag | name != current_name || tag != current_tag)
        }
    }
}
expr failed {
    create(name, tag | name == current_name && tag == current_tag);
    within (>300) {
        zero_or_more {
            one_of {
                ignore_irrelevant
            } or {
                create(name, tag | name == current_name && tag == current_tag)
            }
        };
        one_of {
            create(name, tag)
        } or {
            fetch(name, tag)
        }
    }
}
ignore_any;
failed
\end{lstlisting}

In addition to PTDAs, SyMon supports a high-level specification language defined based on \emph{timed regular expressions}~\cite{DBLP:journals/jacm/AsarinCM02}.
\cref{code:symon-specification} shows an example.
In this expression, two kinds of events (``\texttt{create}'' and ``\texttt{fetch}'') and two parameters over strings (``\texttt{current\_name}'' and ``\texttt{current\_tag}'') are used.
To concisely define the main expression, three subexpressions (``\texttt{ignore\_any}'', ``\texttt{ignore\_irrelevant}'', and ``\texttt{failed}'') are defined.
SyMon constructs a PTDA from a high-level expression and performs monitoring using it.
In our case study, we only use this high-level expression rather than PTDAs.
The details of the high-level expression are omitted.

\section{Case study}
\subsection{Target system}
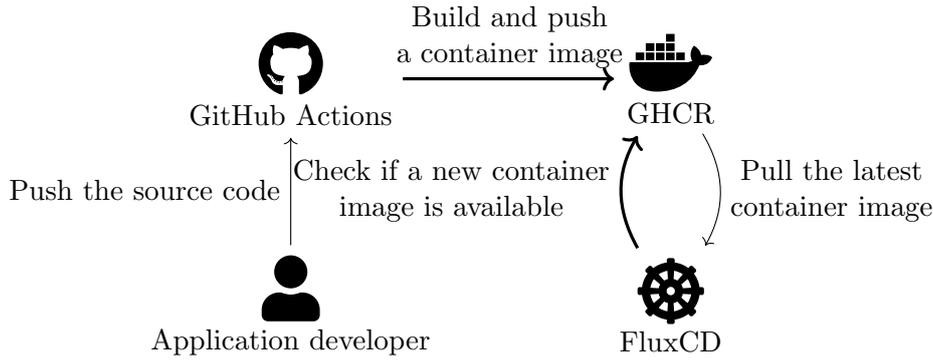
\begin{figure}[tb]
    \centering
    \begin{tikzpicture}[shorten >=1pt,scale=1.00,every node/.style={transform shape}, initial text={}]
        \node (developer) at (0, 0)[align=center] {\Huge \faUser \\ Application developer};
        \node (github) at (0, 3)[align=center] {\Huge \faGithub \\ GitHub Actions};
        \node (docker) at (5, 3)[align=center] {\Huge \faDocker \\ GHCR};
        \node (app) at (5, 0)[align=center] {\Huge \faDharmachakra \\ FluxCD};
        \path[->]
            (developer) edge node[left, align=center] {Push the source code} (github)
            (docker) edge[bend left] node[right, align=center] {Pull the latest\\ container image} (app)
        ;
        \path[->, very thick]
            (github) edge node[above, align=center] {Build and push\\ a container image} (docker)
            (app) edge[bend left] node[left, align=center] {Check if a new container\\ image is available} (docker)
        ;
    \end{tikzpicture}
    \caption{Outline of the target system. Events observed by the monitor correspond to the actions shown by thick arrows.}\label{figure:target_system}
\end{figure}

We monitored a CD system shown in \cref{figure:target_system}. The system consists of GitHub Actions, GitHub Container Registry (GHCR), FluxCD, and an application running on Kubernetes. Application developers first push the application's source code to GitHub. GitHub Actions is then triggered automatically to build an image based on the source code and push it to GHCR. FluxCD regularly polls the image registry, and when it finds a new image, it updates the image of the application running on Kubernetes.

The logs output by the CD system components were used to confirm that the expected time constraints were met. Two types of logs were collected: the first is the Building Phase log, which indicates that an image was pushed to GHCR; the second is the FluxCD polling log, which indicates that FluxCD accessed GHCR and resolved the latest image tag. Throughout the paper, we label the former a ``\texttt{create}'' event and the latter a ``\texttt{fetch}'' event. In this case study, we regard a deployment update as \emph{detected} when FluxCD outputs a ``\texttt{fetch}'' event. Thus, our property checks the delay from an image push (``\texttt{create}'') to FluxCD's detection of the new tag (``\texttt{fetch}'').

We collected the first log as follows. When an image is pushed to GHCR, GitHub's webhook feature notifies an external API server of the event. We created a Registry Monitor to receive the webhooks. When it receives a webhook request, the Registry Monitor prints to standard output a log of the images that were pushed. The log is output in JSON format and includes the package name and tag name as shown in \cref{code:image-pushed-log}.

\begin{lstlisting}[caption=Log indicating that an image has been pushed,label=code:image-pushed-log,style=json,float]
{
  "time":"2025-07-02T02:50:46.42462649Z",
  "package_name":"auth-frontend",
  "package_tag":
    "stg-9c8f5e28c2c7d78da2648f5eaa62216038cbd1fd-1458"
  ....
}
\end{lstlisting}

The second log is obtained from FluxCD. FluxCD periodically polls the image registry and outputs information about the latest images. The log is output in JSON format as shown in \cref{code:flux-polling-log}, with the package and tag names embedded in the \texttt{msg} field.

\begin{lstlisting}[caption=Log of FluxCD polling GHCR,label=code:flux-polling-log,style=json,float]
{
  "level":"info",
  "ts":"2025-07-03T07:06:59.990Z",
  "msg":"Latest image tag for
   ghcr.io/piny940/auth-frontend resolved to
   stg-9c8f5e28c2c7d78da2648f5eaa62216038cbd1fd-1458...",
...
}
\end{lstlisting}

We preprocessed these logs and converted them to a format compatible with SyMon. The logs after preprocessing are shown in \cref{code:log-for-symon}.
Specifically, we performed the following three preprocessing steps.

\begin{enumerate}
    \item Add labels indicating the log type: ``\texttt{create}'' for logs indicating that an image has been pushed and ``\texttt{fetch}'' for logs of FluxCD polling GHCR.
    \item Extract the package name and tag name from each log.
    \item Convert timestamps to Unix time.
\end{enumerate}

\subsection{Specification}\label{subsec:specification}

We tested whether FluxCD detects image updates within five minutes and within ten minutes after they are pushed to GHCR. We used SyMon's specification shown in \cref{code:symon-specification}. The specification reports a violation when a ``\texttt{create}'' event for a given name and tag is not followed by a matching ``\texttt{fetch}'' event within a timeout (\qty{300}{\second} for five minutes or \qty{600}{\second} for ten minutes), ignoring unrelated events.

Lines 2--5 declare variables representing the name and tag to be monitored. Only events that match these values are focused on. Lines 6--13 define \texttt{create} and \texttt{fetch} as log events. The \texttt{ignore\_any} subexpression defined in lines 14--22 matches any number of \texttt{create} or \texttt{fetch} events; this subexpression ignores harmless parts of the event sequence. Lines 23--31 define the \texttt{ignore\_irrelevant} subexpression, which ignores \texttt{create} and \texttt{fetch} events that do not match \texttt{current\_name} or \texttt{current\_tag} as irrelevant events. Lines 32--48 define the \texttt{failed} subexpression, which is the main body of this specification, i.e., the anomaly to be detected. The content is represented in the following three steps.
\begin{enumerate}
    \item A \texttt{create} event occurs (l33).
    \item More than the chosen timeout (either \qty{300}{\second} or \qty{600}{\second}) passes without observing a matching \texttt{fetch} event, while only irrelevant events and possibly repeated matching \texttt{create} events occur (l34--l41).
    \item Some event occurs (l42--l46).
\end{enumerate}

Lines 49--50 specify that the monitor searches for occurrences of \texttt{failed} after \texttt{ignore\_any}.

\subsection{Results and Discussion}

We checked whether the system's behavior satisfied the specification in \cref{subsec:specification}. We analyzed logs from five days (12,758 entries), including 12 ``\texttt{create}'' events.
The logs used for benchmarking are available on GitHub~\cite{symon-log}.

With the 5-minute threshold, SyMon reported five outputs, as shown in \cref{code:symon-result}, indicating that FluxCD did not always detect the new image tag within five minutes after the image was pushed. Note that FluxCD outputs \texttt{fetch} logs periodically; thus SyMon can report multiple outputs for the same \texttt{create} event. In contrast, with the 10-minute threshold, SyMon reported no outputs. This means that FluxCD detected the new tag within ten minutes in the collected logs.

\begin{lstlisting}[caption=output of SyMon,label=code:symon-result,style=symon_result,float]
@1751425023.000000.     (time-point 9443)       x0 == auth-frontend     x1 == stg-9c8f5e28c2c7d78da2648f5eaa62216038cbd1fd-1458 true
@1751425023.000000.     (time-point 9443)       x0 == auth-example      x1 == stg-9c8f5e28c2c7d78da2648f5eaa62216038cbd1fd-1458 true
@1751425050.000000.     (time-point 9444)       x0 == auth-example      x1 == stg-9c8f5e28c2c7d78da2648f5eaa62216038cbd1fd-1458 true
@1751425050.000000.     (time-point 9444)       x0 == auth-frontend     x1 == stg-9c8f5e28c2c7d78da2648f5eaa62216038cbd1fd-1458 true
@1751425052.000000.     (time-point 9445)       x0 == auth-example      x1 == stg-9c8f5e28c2c7d78da2648f5eaa62216038cbd1fd-1458 true
\end{lstlisting}

We also measured the execution time of SyMon against 5, 10, and 15 days of logs. We ran the test on a computer with an AMD Ryzen 5 5600 (3.5\,GHz) CPU and 16\,GB of memory. The time taken to execute SyMon is shown in \cref{tab:execution-time}.
It took 360 milliseconds to check 12,758 log entries from five days.
For comparison, we ran tests on 25,223 log entries from 10 days and 41,151 log entries from 15 days, and it finished in 376 and 398 milliseconds, respectively. This indicates that execution time remains sub-second even as the number of log entries increases.
Since the monitored time bounds are on the order of minutes (\qtyrange{300}{600}{\second}), these sub-second execution times suggest that the monitoring overhead is small in our setting.

\begin{table}[tb]
    \centering
    \caption{Time taken to execute SyMon}
    \begin{tabular}{c@{\hspace{15pt}}c@{\hspace{15pt}}c} \toprule
        Number of Days & Number of Entries & Execution Time (ms) \\ \midrule
        5 & 12,758 & 360 \\
        10 & 25,223 & 376 \\
        15 & 41,151 & 398 \\ \bottomrule
    \end{tabular}
    \label{tab:execution-time}
\end{table}

\section{Conclusions and Perspectives}
In this paper, we used SyMon to confirm that a FluxCD-based CD system operates according to the specification. The event logs showed that FluxCD did not always detect a new image within five minutes after it was pushed to GHCR, whereas it always did so within ten minutes in the collected logs.

One area for improvement in SyMon is support for JSON-formatted logs without preprocessing. Currently, SyMon cannot handle JSON-formatted logs directly, so it is necessary to preprocess the logs into a format that SyMon can handle. However, describing preprocessing individually for each system is a significant implementation burden. We plan to further improve SyMon to build a more practical monitoring system.

\subsubsection*{Acknowledgements}

This work is partially supported by
    JSPS KAKENHI Grant No.~22K17873,
    JST BOOST Grant No.~JPMJBY24H8, and
    JST PRESTO Grant No.~JPMJPR22CA.

\bibliography{k8s-symon}
\bibliographystyle{splncs04}%

\end{document}